\documentclass{appolb}
\usepackage{graphicx}
\usepackage{amsmath}
\usepackage{amssymb}
\usepackage{bm}
\usepackage{color}
\usepackage{hyperref}

\begin{document}
\title{Thermalization of weakly coupled non-Abelian plasmas at next-to-leading order%
\thanks{Presented at Quark Matter 2022}
}
\author{ Yu Fu\footnote{Speaker.}
\address{Key Laboratory of Quark and Lepton Physics (MOE) \& Institute of Particle Physics, Central China Normal University, Wuhan 430079, China}
\\[3mm]
{Jacopo Ghiglieri 
\address{SUBATECH, Universit\'e de Nantes, IMT Atlantique, IN2P3/CNRS}
}
\\[3mm]
{Shahin Iqbal
\address{National Centre for Physics, Quaid-i-Azam University Islamabad, Pakistan}
}
\\[3mm]
Aleksi Kurkela 
\address{Faculty of Science and Technology, University of Stavanger, 4036 Stavanger, Norway}
}
\maketitle
\begin{abstract}
We employ the QCD kinetic theory, including next-to-leading(NLO) order corrections in coupling constant, to study the evolution of weakly coupled non-Abelian plasmas towards thermal equilibrium. For two characteristic far-from-equilibrium systems with either under- or over-occupied initial conditions, the NLO corrections remain well under control for a wide range of couplings, and the overall effect of NLO corrections is a reduction in the time required for thermalization.
\end{abstract}
  
\section{Introduction}

In the early stage of ultra-relativistic heavy-ion collisions, a far-from-equilibrium system of quarks and gluons is created initially.  One of the central goals in the study of heavy-ion physics is to establish a description of the approach to thermal equilibrium. A \textit{bottom-up thermalization} scenario is proposed in which the equilibration process is considered in the weak coupling limit\cite{Baier:2000sb}. In the initial state of the system, the energy density is carried by a large number of over-occupied low-energy modes of excitation $f\gg1$. This system subsequently evolves to a collection of under-occupied states, $f\ll1$, at high momentum  $p\gg T$ before reaching thermal equilibrium. Within the framework of effective kinetic theory(EKT) at leading order(LO) in coupling established by Arnold, Moore, Yaffe(AMY)\cite{Arnold:2002zm}, the bottom-up mechanism has been confirmed by numerically solving 2+1D kinetic equation of pure gluon plasma under longitudinal expansion at small t'Hooft coupling ($\lambda=g^2N_c$) limit \cite{Kurkela:2015qoa}. However, one should also notice that when extrapolating to moderate values of coupling, e.g., $\lambda=10$, such a bottom-up scenario becomes less obvious. 

In the context of weak-coupling kinetic theory, one extrapolate the results at weak-coupling limit to ``realistic'' coupling for phenomenological applications in heavy-ion collisions. Therefore, it is important to improve the accuracy and to test the validity and robustness of the weak-coupling expansion by finding the first subleading corrections to the weak-coupling results. When extrapolating to the ``realistic'' coupling, the NLO calculations for some (near-)equilibrium quantities have shown that, for example, NLO corrections to the heavy quark diffusion coefficient completely overtake LO results \cite{Caron-Huot:2007rwy} and the NLO result for shear viscosity $\eta/s$ is much smaller than LO result\cite{Ghiglieri:2018dib}. Given the notoriously poor convergence of perturbative theory at finite temperature as shown in calculations for these (near-)equilibrium quantities, one also naturally questions if it is allowed to extrapolate the results in weak-coupling limit to a moderate value of coupling around $\lambda=10$ for far-from-equilibrium system. In our recent work \cite{Fu:2021jhl}, we present the first weak-coupling description of thermalization of pure Yang-Mills plasmas from isotropic initial conditions at NLO accuracy by numerically solving the QCD effective kinetic theory.

\section{Effective kinetic theory beyond leading order}
In the weak coupling limit, the EKT can describe the time evolution of systems where the typical occupancies of gluons are purterbative $f(p)\ll1/\lambda$ and have typical momenta larger than the in-medium screening scale $p\gg m^2\equiv 4\lambda \int_{\bf p }f(p)/p$. At leading order in $\lambda$, the theory is defined through the effective Boltzmann equation for color averaged distribution of gluons\cite{Arnold:2002zm},
 \begin{equation}
 \partial_t f(p,t) = -\mathcal{C}_{2\leftrightarrow 2}[f](p)-\mathcal{C}_{1\leftrightarrow 2}[f](p).
 \label{locoll}
 \end{equation}
Here, we consider only isotopic pure gluon systems. Isotropy guarantees that there is no plasma instability. $\mathcal{C}_{2\leftrightarrow 2}[f](p)$ represents the collision integral for leading order elastic $2\leftrightarrow 2$ scattering. It depends on effective matrix elements $|\mathcal{M}|^2$, in which the soft divergence in $t$ and $u$ channels is regulated by the isotropic screening at the scale $m$. $\mathcal{C}_{1\leftrightarrow 2}[f](p)$ accounts for the effective $1\leftrightarrow 2$ slitting process with splitting rate $\gamma$. The splitting rate depends on $m$ and effective temperature $T_* = \frac{2\lambda}{m^2}\int_{\mathbf{p}}f_p(1+f_p)$, and is 
obtained by resumming multiple interactions with the medium and includes the Landau-Pomeranchuk-Migdal(LPM) suppression of collinear radiation. The precise form of the collision operators at LO is given in \cite{Kurkela:2014tea}.

The physical picture of EKT has been extended to NLO accuracy for the case where one follows the evolution of a dilute set of hard particles interacting with a thermal medium\cite{Ghiglieri:2015ala}. These NLO corrections come from the self-interactions of thermal soft gluons. In finite temperature perturbative calculations, the statistical function of these soft modes results in relative $\mathcal{O}(\lambda^{1/2})$ NLO corrections, rather than $\mathcal{O}(\lambda)$ in vacuum perturbative theory.
These NLO corrections can be also extended to some far-from-equilibrium systems in which $f(m\lesssim p\ll T_*)\approx T_*/p$. For such systems, these NLO corrections are $\mathcal{O}(\lambda T_*/m)$, which arises from multiplying the perturbative expansion factor $\lambda$ for gluon loops by occupation number of soft gluon $f(m)\approx T_*/m$.

The detailed implementation of NLO corrections can be found in \cite{Fu:2021jhl} and we briefly summarize these NLO corrections as follows: 
\begin{enumerate}
\item The splitting rate $\gamma(m,T)$ is modified by considering the LPM suppression including two $\mathcal{O}(\lambda T_*/m)$ corrections: the LO effective gluon  mass squared $m_g^2$ gets shifted to $m_g^2{}_\mathrm{NLO}=m_g^2+\delta m_g^2$ and the soft LO scattering kernel $C(b)$ is modified as $C_\mathrm{NLO}(b)=C(b)+\delta C(b)$. 
\item A semi-collinear $1\leftrightarrow 2$ splitting rate $\gamma|_{semi}$ encodes the splitting rate at wider-angle $1\leftrightarrow2$ processes in a new kinematical region, as well as contributions to longitudinal momentum diffusion arising from soft legs in $1\leftrightarrow 2$ processes and from soft loops in $2\leftrightarrow2$ processes. 
\item An $\mathcal{O}(\lambda T_*/m)$ correction contained in the original LO $2\leftrightarrow 2$ collision kernel is subtracted to avoid double countings. 
\end{enumerate}
In addition, in kinetic theory, it is possible to construct collision operators that are equivalent up to a given order but differ by subleading corrections. Given this property, we develop two different schemes, \emph{scheme 1} and \emph{scheme 2}, to implement the NLO corrections to the effective gluon mass and the soft scattering kernel, which affect the calculation of splitting rate $\gamma(m,T)$ at NLO. In scheme 1, we treat corrections to effective gluon mass and scattering kernel as perturbations to LO. In the not-strict implementation, scheme 2, these corrections are not purterbation to LO, including partial resummation of higher order effects. 
These schemes formally differ at higher order, and their difference can be taken as a measure of the uncertainty in the NLO calculation.

\section{Results and analysis}
Based on the EKT corrected to NLO accuracy, we follow the time evolution of two typical far-from-equilibrium systems, pure Yang-Mills plasmas with under- and over-occupied initial distribution. For the under-occupied initial condition we will use $ f(p) = A \exp(-\frac{(p-Q)^2}{(Q/10)^2}) +  n_B(p, T_{\rm init})$, where $A \approx (0.419 Q/T)^{-4}$ and $T_{\rm init} \approx 0.562T$. $n_B$ is the Bose--Einstein distribution. Such a distribution mimics the situation in the last stage of bottom-up thermalization. In the over-occupied case, the system is initialized with the scaling solution  $f(p)= (Q t)^{-4/7}\lambda^{-1} \tilde f(\tilde p)$, where $ \tilde p\equiv (p/Q)(Q t)^{-1/7}   $ and $\tilde f(\tilde p) \equiv (0.22 e^{-13.3 \tilde p}+2.0e^{-0.92 \tilde{p}^2})/\tilde p$. We define the effective temperatures as $T_\alpha = (\frac{2\pi^2}{\Gamma(\alpha+3)\zeta(\alpha+3)} \int \frac{d^3 p}{(2\pi)^3} p^\alpha f(p))^\frac{1}{\alpha + 3}$. The thermalization time is determined by $\left(T_0(t_{\rm eq})/T_1(t_{\rm eq}) \right)^{\pm4} = 0.9$, where we use "$+$" and "$-$" for under- and over-occupied systems, respectively.
\begin{figure*}[ht]
    \centering
    \includegraphics[width=\textwidth]{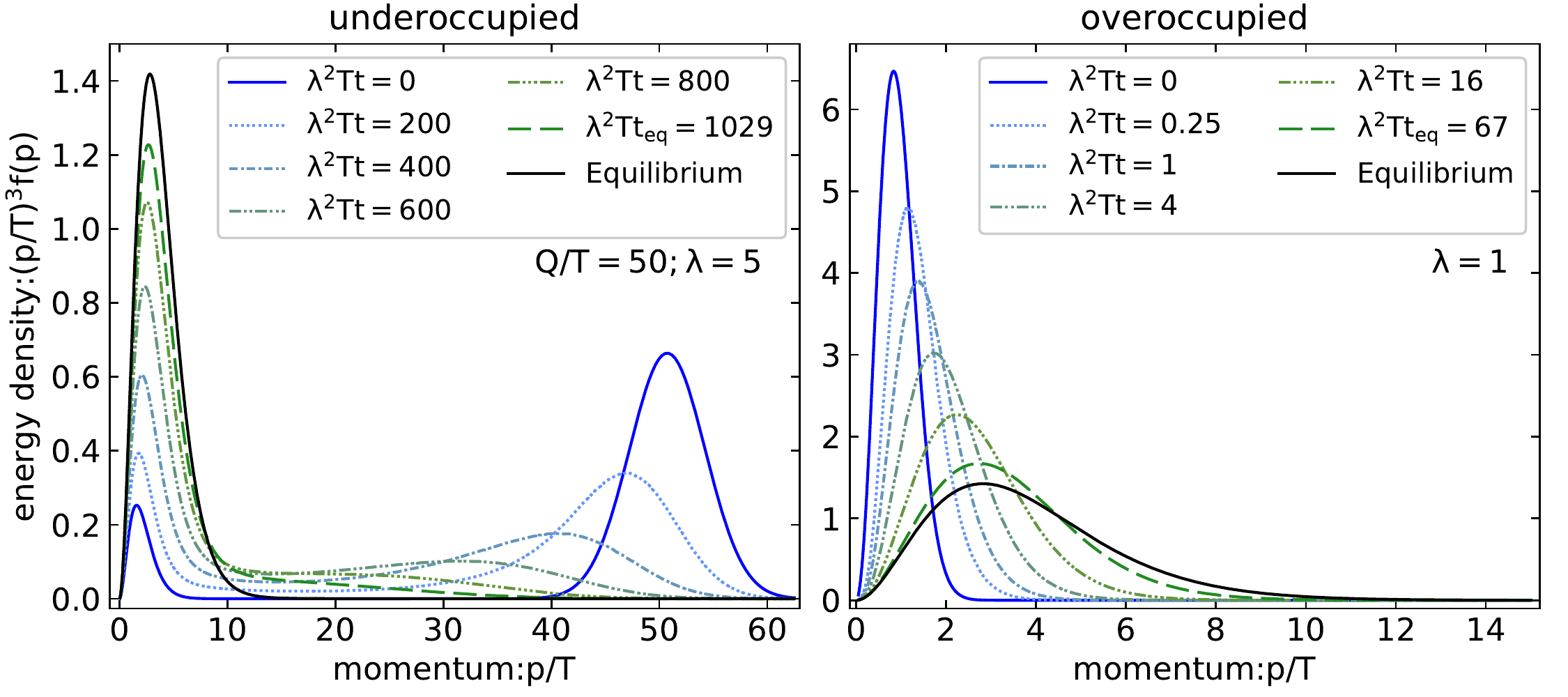}
    \caption{Time evolution from the initial conditions 
    in solid blue lines to the final equilibrium state in solid black. The dotted and dashed lines show intermediate steps upon solving the NLO kinetic theory (scheme 2). The values of the couplings
    are $\lambda=5$ and $\lambda=1$ respectively.
     \label{fig:timeevo}}
\end{figure*}
\begin{figure*}[ht]
    \begin{center}
        \includegraphics[width=\textwidth]{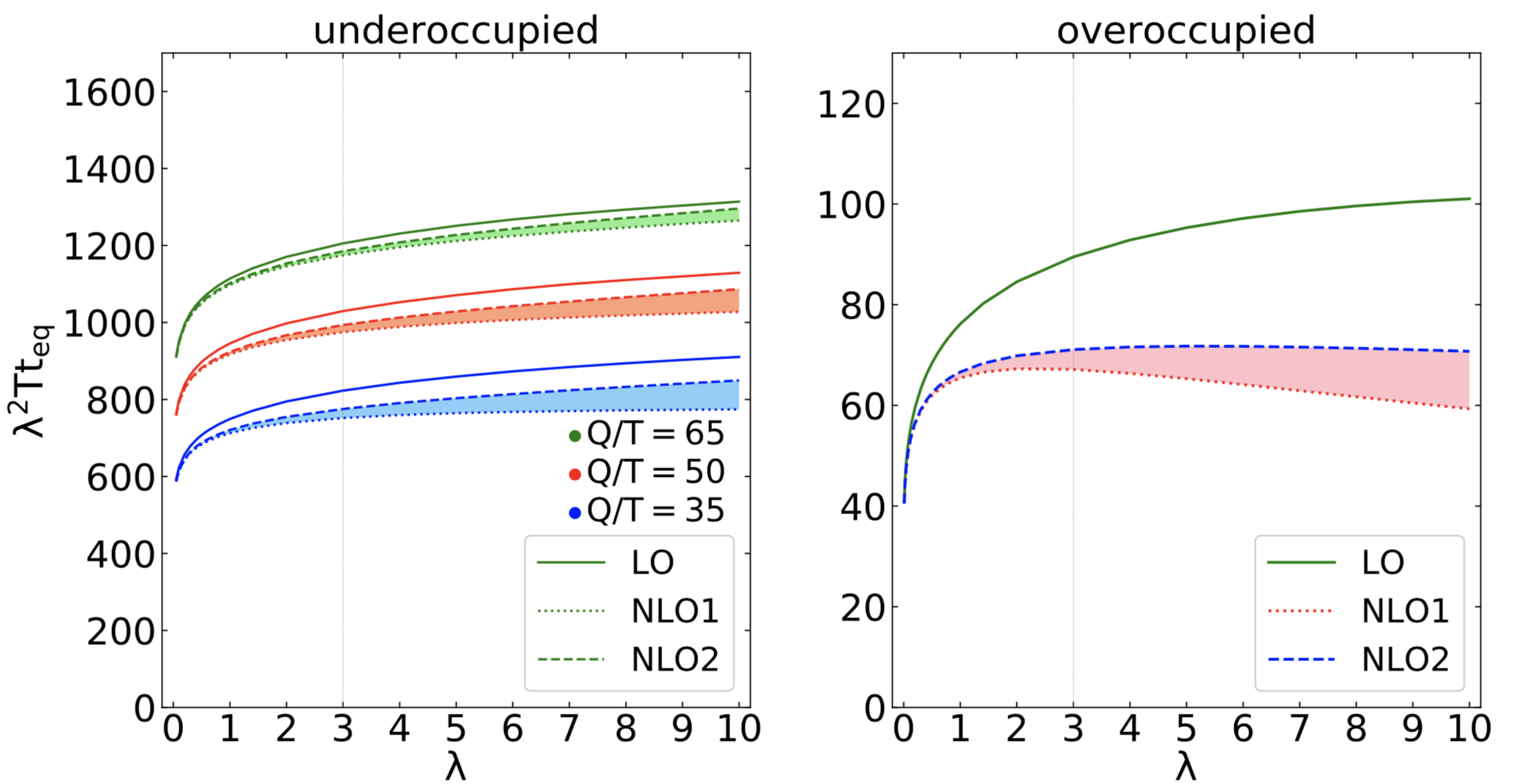}
    \end{center}
  \caption{Thermalization times as a function of the coupling.   The shaded band between the two NLO schemes can be taken as a first indication of the theory uncertainty. The coupling $\lambda=3$ for which $m=T$ in thermal equilibrium is indicated by a vertical gray line.
      \label{fig_teq} 
   }
\end{figure*}
Shown in Fig.~\ref{fig:timeevo} are the NLO (scheme 2) thermalization processes of under-occupied system with $Q=50$ and $\lambda = 5$ (left panel) and of over-occupied system with $\lambda = 5$ (right panel). The green dashed lines correspond to the states that satisfy the conditions for determining thermalization time. The NLO evolutions of these systems exhibit the same qualitative features as their LO counterparts.

Varying the values of initial characteristic momentum $Q$ for under-occupied system and coupling constant $\lambda$ for both systems, we determine the thermalization times and investigate the dependence on $\lambda$ (and on $Q$ for under-occupied case) at LO and NLO (scheme 1 and scheme 2). The relation between thermalization times and coupling constants for both systems are presented in Fig.~\ref{fig_teq}. We find that the LO thermalization time is well descried for $\lambda < 5$ by a fit $ \lambda^2 T t_{eq}^\mathrm{LO} \approx  (Q/T)^{1/2} (173. + 9.8 \log\lambda)-277$. For $\lambda < 1$ and $20 < Q < 80$ the NLO correction in both schemes is approximately given by $t^\mathrm{LO}_{eq}/t_{eq}^\mathrm{NLO}  \approx 1 + \lambda^{1/2}\left( 0.22-0.05 \log\left(Q/T\right)\right)$.
Similarly, for the overoccupied case, we obtain ${\lambda^2 T t_{eq}^\mathrm{LO}} \approx 76./(1 - 0.19 \log{\lambda})$ and $t^\mathrm{LO}_{eq}/t_{eq}^\mathrm{NLO}  \approx 1 + 0.14 \lambda^{1/2}$.

When $\lambda \lesssim 3$, NLO corrections constitute merely a $5\%$($20\%$) reduction of the $t_{eq}$ in the under-occupied(over-occupied) cases. When $\lambda\to 0$, we observe that the difference between the two NLO schemes vanishes faster than their difference to LO for both systems, meaning that the observed differences from the LO are true NLO corrections and are not contaminated by the scheme differences that affect the result beyond the NLO accuracy. 

When $3 \lesssim \lambda\lesssim 10$, in the under-occupied case the difference between the two NLO schemes becomes 
comparable to the size of the NLO correction itself, which
shows theoretical uncertainty arising from corrections beyond NLO. However, we also observe that the corrections remain below 10\%-level even for these large value of the coupling. In the over-occupied case the correction reaches 40\%-level, with only a moderate spread between the two schemes.

\section{Conclusion}
In this work we provide a NLO weak-coupling description of the thermalization process of far-from-equilibrium  (under- and over-occupied) pure Yang-Mills isotropic plasmas. From our numerical results, we conclude that the NLO evolutions of these systems exhibit the same qualitative evolution patterns as their LO counterparts and the overall effect of NLO corrections is to reduce the time needed to reach thermal equilibrium. Furthermore, in contrast to NLO calculations for some quantities, e.g. heavy quark diffusion coefficient and shear viscosity, which suffer from poor convergence and show remarkable theoretical uncertainties, soft corrections in the study of isotropic thermalization, however, are well under control for a wide range of couplings at NLO accuracy. 

\section{Acknowledgement}
J.G. acknowledges support by a PULSAR grant from the R\'egion Pays de la Loire. S.I. and Y.F. were supported in part by the National Natural Science Foundation of China under Grant Nos. 11935007, 11221504, 11890714 and 11861131009.


\begin{thebibliography}{99}

\bibitem{Baier:2000sb}
R. Baier, A.H. Mueller, D. Schiff, and D.T. Son,
```Bottom up' thermalization in heavy ion collisions,''
\textit{Phys. Lett. B}, 502:51–58, 2001.


\bibitem{Arnold:2002zm}
P. Arnold, G. Moore, and L. Yaffe,
``Effective kinetic theory for high temperature gauge theories,''
\textit{JHEP}, 01:030, 2003.


\bibitem{Kurkela:2015qoa}
A. Kurkela and Y. Zhu,
``Isotropization and hydrodynamization in weakly coupled heavy-ion collisions,''
\textit{Phys. Rev. Lett.}, 115(18):182301, 2015.



\bibitem{Caron-Huot:2007rwy}
S. Caron-Huot,and G. Moore,
``Heavy quark diffusion in perturbative QCD at next-to-leading order,''
\textit{Phys. Rev. Lett.}, 100:052301, 2008.


\bibitem{Ghiglieri:2018dib}
J. Ghiglieri, G. Moore, and D. Teaney,
``QCD Shear Viscosity at (almost) NLO,''
\textit{JHEP}, 03:179, 2018.


\bibitem{Fu:2021jhl}
Y. Fu, J. Ghiglieri S. Iqbal and A. Kurkela,
``Thermalization of non-Abelian gauge theories at next-to-leading order,''
\textit{Phys. Rev. D},105(5):054031, 2022.

\bibitem{Kurkela:2014tea}
A. Kurkela and E. Lu,
``Approach to Equilibrium in Weakly Coupled Non-Abelian Plasmas,''
\textit{Phys. Rev. Lett.}, 113(18):182301, 2014.


\bibitem{Ghiglieri:2015ala}
J. Ghiglieri, G. Moore, and D. Teaney,
``Jet-Medium Interactions at NLO in a Weakly-Coupled Quark-Gluon Plasma,''
\textit{JHEP}, 03:095, 2016.


\end{thebibliography}
\end{document}